\documentclass[%
 aip,
 amsmath,amssymb
 reprint,%
]{revtex4-1}

\usepackage{graphicx}% Include figure files
\usepackage{bm}% bold math
\usepackage[utf8]{inputenc}
\usepackage[T1]{fontenc}
\usepackage{mathptmx}
\usepackage{etoolbox}

\usepackage{dsfont}
\usepackage{amssymb}
\usepackage{tipa}
\usepackage{stmaryrd}
\usepackage{amsmath}
\usepackage{bm}
\usepackage{yfonts}

\makeatletter
\def\@email#1#2{%
 \endgroup
 \patchcmd{\titleblock@produce}
  {\frontmatter@RRAPformat}
  {\frontmatter@RRAPformat{\produce@RRAP{*#1\href{mailto:#2}{#2}}}\frontmatter@RRAPformat}
  {}{}
}%
\makeatother
\begin{document}

%\preprint{AIP/123-QED}

\title[]{A quantum number theory}
% Force line breaks with \\
\author{L. Daiha}
 \altaffiliation[]{E-mail: lucas.daiha@ufba.br}%Lines break automatically or can be forced with \\

%\author{B. Author}%
% \email{Second.Author@institution.edu.}
%\affiliation{ 
%Authors' institution and/or address%\\This line break forced with %\textbackslash\textbackslash
%}%

\author{R. Rivelino}
 \homepage{E-mail: rivelino@ufba.br}
\affiliation{%
Instituto de F\'{\i}sica, Universidade Federal da Bahia, 40210-340 Salvador, Bahia, Brazil %\\This line break forced% with \\
}%
\date{\today}% It is always \today, today,

\begin{abstract}
We employ an algebraic procedure based on quantum mechanics to propose a `quantum number theory' (QNT) as a possible extension of the `classical number theory'. We built our QNT by defining pure quantum number operators ($q$-numbers) of a Hilbert space that generate classical numbers ($c$-numbers) belonging to discrete Euclidean spaces. To start with this formalism, we define a 2-component natural $q$-number $\textbf{N}$, such that $\mathbf{N}^{2} \equiv N_{1}^{2} + N_{2}^{2}$, satisfying a Heisenberg-Dirac algebra, which allows to generate a set of natural $c$-numbers $n \in \mathbb{N}$. A probabilistic interpretation of QNT is then inferred from this representation. Furthermore, we define a 3-component integer $q$-number $\textbf{Z}$, such that $\mathbf{Z}^{2} \equiv Z_{1}^{2} + Z_{2}^{2} + Z_{3}^{2}$ and obeys a Lie algebra structure. The eigenvalues of each $\textbf{Z}$ component generate a set of classical integers $m \in \mathbb{Z}\cup \frac{1}{2}\mathbb{Z}^{*}$, $\mathbb{Z}^{*} = \mathbb{Z} \setminus \{0\}$, albeit all components do not generate $\mathbb{Z}^3$ simultaneously. We interpret the eigenvectors of the $q$-numbers as `$q$-number state vectors' (QNSV), which form multidimensional orthonormal basis sets useful to describe state-vector superpositions defined here as qu$n$its. To interconnect QNSV of different dimensions, associated to the same $c$-number, we propose a quantum mapping operation to relate distinct Hilbert subspaces, and its structure can generate a subset $W \subseteq \mathbb{Q}^{*}$, the field of non-zero rationals. In the present description, QNT is related to quantum computing theory and allows dealing with nontrivial computations in high dimensions.
\end{abstract}


\maketitle

%\begin{quotation}
%The ``lead paragraph'' is encapsulated with the \LaTeX\ 
%\verb+quotation+ environment and is formatted as a single paragraph before the %first section heading. 
%(The \verb+quotation+ environment reverts to its usual meaning after the first %sectioning command.) 
%Note that numbered references are allowed in the lead paragraph.
%
%The lead paragraph will only be found in an article being prepared for the journal %\textit{Chaos}.
%\end{quotation}

\section{Introduction}

Algebraic structures in quantum mechanics are of two-fold interest, namely for explaining microscopic physical phenomena and for applications in quantum computing/information \cite{haroche2020,luo2019,deville2019,dogra2018,kurzynski2016,caputa2019}. An interesting example of this successful mathematical structure refers to the problem of universal quantum gates \cite{kortryk2016,bertlmann2008,erhard2018}, which is recognized as a particular case of the Lie algebras applied to quantum computers \cite{padmanabhan2020}. A popular and useful representation of this algebraic structure is related to the Pauli matrices, particularly applied to quantum logic gates \cite{nielsen2010} for qubits. Within a pure mathematical framework, the Pauli matrices define a faithful representation of the real Clifford algebra \cite{clifford} $C\ell_{3,0}(\mathbb{R})$ on the two-dimensional complex vector space $\mathbb{C}^{2}$. Thus, along with the corresponding identity matrix, the Pauli matrices form an orthogonal basis, in the sense of Hilbert-Schmidt operators \cite{hilbert}; for the real Hilbert space of 2$\times$2 complex Hermitian matrices, or for the complex Hilbert space of all 2$\times$2 matrices. Extending this algebraic formalism, one obtains generalized matrices for higher dimensions \cite{stephany1979}, as in the case of the Gell-Mann matrices \cite{wang2017}, which constitute a proper representation for qutrits and are also useful in quantum computing applications \cite{scirep2013}.

As a consequence of this abstract algebraic structure of quantum mechanics, from which emerges a discrete pattern of energy levels, there is also an interplay with classical number theory \cite{friedmann2015,cotti2020,gleisberg2018,ramos2014}. In this sense, a number-theoretical approach has been proposed to compute the degeneracy spectrum of black holes in the context of loop quantum gravity \cite{agullo2018}. More recently, number-theoretical quantum states have been proposed in terms of qubits \cite{sierra2020} to encode arithmetic properties of the primes. Famously, number theory is mainly concerned with the properties of the integers, which are also the basis for a quantum computer \cite{feynman1982,shor1997}. Hence, there is really a great motivation to extend number theory for a suitable quantum structure. In 2004, Tokuo \cite{tokuo2004} developed a number theory within the framework of a proposed typed quantum logic \cite{tokuo2003}. However, the problem with this theory is exactly in establishing a quantum logic with an adequate set of axioms to derive the fundamental theorems.

In this proposal, we exploit the algebraic structure of quantum mechanics to construct a representation for a possible quantum number theory (QNT). First, we define a 2-component natural $q$-number (in the Dirac's prescription \cite{dirac}) in the form $\textbf{N}^2 \equiv N_{1}^{2} + N_{2}^{2}$, satisfying the commutation relation $[N_{1}, N_{2}] = iI$. This definition allows to obtain a set of `classical' natural numbers. Second, we define a 3-component integer $q$-number in the form $\mathbf{Z}^{2} \equiv Z_{1}^{2} + Z_{2}^{2} + Z_{3}^{2}$, such that $[Z_{i}, Z_{j}] = i\epsilon_{ijk}Z_{k}$ and [$\mathbf{Z}^{2},Z_{p}$] = 0 for all $p$ indices. From the eigenvalue problems of $\mathbf{Z}^{2}$ and $Z_{p}$, we find that the solutions are polynomials with whole coefficients for each odd dimension, which enable to generate a set of `classical' integers $\mathbb{Z}$. Hence, in QNT we obtain the $c$-numbers by systematically solving characteristic eigenvalue problems related to the operators $\textbf{N}$ and $\textbf{Z}$. We show that QNT is also related to quantum computing theory in higher dimensions than the dimensions of qubits and qutrits \cite{scirep2019}. Within an algebraic framework, the present QNT can be viewed as an interesting and general description to broaden the meaning of number. Overall, this leads to the novel concepts of `number state' and `basic unit of number information' in number theory. 

This paper is organized as follows. In Sec. II, we define a natural $q$-number that leads to a native even-odd partition of the Hilbert space, generalizing the concept of parity in number theory. We also define a qu$n$it in the natural basis, which is quantum-statistically interpreted, free of the ontological measurement problem common in quantum mechanics. A general qu$n$it in this basis implies in Poisson and quasi-Poisson distributions of even and odd numbers, respectively. Section III introduces a 3D integer representation, preserving a Lie algebra that gives rise to algebraic numbers from eigenvalue problems of its components. In Sec. IV, we propose a quantum mapping, in the sense of QNT, to interconnect states of integer $c$-numbers with different dimensions, resulting from the accidental degeneracy of $\textbf{Z}$. Furthermore, such a construction leads to an algebraic description of finding the set of rationals. Finally, we present in Sec. V a generalization of a Z-representation related to the SU($n$) group and give our concluding remarks in Sec. VI.

\section{The Natural Number Representation}

To construct the algebraic formalism of our proposed QNT, we define pure `quantum number operators', in the sense of Hilbert-Schmidt operators acting on a non-commutative configuration space $\mathcal{H}$. We make use of the $q$-number concept within the Dirac's prescription \cite{dirac}, obeying well-established quantum algebras. Hence, we introduce a two-component natural $q$-number $\mathbf{N}$, whose components satisfy the Heisenberg-Dirac algebra, which also span a Lie algebra, and enable us to generate $\mathbb{N} = \big\{0,1,2,3,...\big\}$. In this construction, the $c$-numbers are obtained from eigenvalue problems related to these operators, whereas the abstract concept of $q$-number eigenvectors increases the number information and, consequently, generalizes the own meaning of number. 

\subsection{A Pythagorean construction of the natural $q$-number}
Let us first consider the properties of the number operator $\mathbf{N}$.

\textbf{Axiom 1.}
Every classical natural number (namely, a natural $c$-number) is an eigenvalue of a natural quantum number operator (namely, a natural $q$-number).

\textbf{Definition 1.}
Let $\mathbf{N} \equiv (N_{1}, N_{2})$ be a two-dimensional (2D) representation of a natural $q$-number satisfying the following properties:
\begin{equation} \label{property1}
    \mathbf{N}^{2} = N_{1}^{2} + N_{2}^{2}
\end{equation}
\begin{equation} \label{property2}
    \mathbf{N} = \mathbf{N}^{\dagger}
\end{equation}
\begin{equation} \label{property3}
    [N_{1}, N_{2}] = iI
\end{equation}
Property (\ref{property1}) is chosen to define a Pythagorean representation of $\mathbf{N}$ in $\mathcal{H}$; property (\ref{property2}) is used to ensure an Hermitian representation of $\mathbf{N}$; and property (\ref{property3}) is required to respect a type of Heisenberg-Dirac algebra with $I$ being the identity operator. In this sense, the number components $N_{1}$ and $N_{2}$ are Hermitian operators in $\mathcal{H}$ or auxiliary $q$-numbers in QNT.

From the components and the properties of the $q$-number $\mathbf{N}$, we introduce the non-Hermitian ladders operators, which allow the factorization of $\mathbf{N}^2$, i.e.,
\begin{equation} \label{nplus}
    N_{+} \equiv N_{1} + iN_{2}
\end{equation}
\begin{equation} \label{nminus}
    N_{-} \equiv N_{1} - iN_{2}
\end{equation}
In this way, property (\ref{property1}) leads to 
\begin{equation} \label{n2D}
    N_{-}N_{+}=\mathbf{N}^{2} - I
\end{equation}

\textbf{Definition 2.}
Let $N \equiv N_{-}N_{+}$ and $N^{\star} \equiv \mathbf{N}^{2}$ be linear natural $q$-numbers, such that $N = N^{\star} - I$ and satisfies the eigenvalue equation 
\begin{equation} \label{nk}
N|n\rangle = n|n\rangle
\end{equation}
with $n \in \mathbb{N}$ and defining an infinite orthonormal basis set $\big\{|0\rangle, |1\rangle, |2\rangle,...\big\}$ of `$q$-number state vectors' (QNSV), denoted in the Dirac's notation \cite{dirac2} of kets ($| \hspace{0.2cm} \rangle$) and corresponding bras ($ \langle \hspace{0.2cm}|$) in the dual space. 

Note that $N$ is positive semidefinite, i.e., $\forall$ $|n\rangle \neq 0$, $\langle n |N| n \rangle \geq 0$ and the eigenvalues of $N$ are non-negative ($\textup{Tr}(N) \geq 0$). Since $N^{\star}$ is a linear function of $N$, both can be diagonalized simultaneously. Then, because of property (\ref{n2D}) we also write the eigenvalue equation
\begin{equation} \label{nsquare}
N^{\star}|n\rangle = (n+1)|n\rangle
\end{equation}
with $N^{\star}$ being a positive definite $q$-number, i.e., it is the generator of $\mathbb{N}^* = \big\{1,2,3,...\big\}$.
We note that the QNSV of $N$ and $N^{\star}$ are the same and have infinite dimensions. However, to compute properties, it is also possible to use a finite matrix representation for them in a Hilbert subspace. Using this algebraic description, we obtain ${N}^{\star}|0\rangle = 1|0\rangle$, where $|0\rangle$ is not a null QNSV and the natural $c$-number 1 can be defined as $\langle 0 |{N}^{\star}| 0 \rangle = 1$. 

For an arbitrary eigenvector $|n\rangle$ of $N$, the square of the norm of $N_{+}|n\rangle$ is positive or null:
\begin{equation} \label{normnat}
    ||N_{+}|n\rangle||^2 = \langle n |N_{-}N_{+}| n \rangle = \langle n |N| n \rangle \geq 0
\end{equation}
Using Eqs. (\ref{nplus}) and (\ref{nminus}), it is also straightforward  to demonstrate that
\begin{equation} \label{Ncomm}
    [N, N_{+}] = -2N_{+},  \quad [N, N_{-}] = 2N_{-}
\end{equation}
From this result, we find that $N_{+}| n \rangle$ and $N_{-}| n \rangle$ are also eigenkets of $N$ associated to the eigenvalues $(n-2)$ and $(n+2)$, respectively. This also implies that $N_{+}| n \rangle$ and $|n-2 \rangle$, with $n \geq 2$, are related up to a multiplicative constant, or equivalently
\begin{equation} 
N_{+}|n+2\rangle = c^{+}_{n}|n\rangle
\end{equation}
Since we are considering orthonormal QNSV, then 
\begin{equation}
\langle n+2 |N_{-}N_{+}| n+2 \rangle = |c^{+}_{n}|^2 
\end{equation}
By convention, we choose $c^{+}_{n}$ as a real and positive $c$-number, such that
\begin{equation} \label{Nminus}
N_{+}|n\rangle = \sqrt{n}|n-2\rangle
\end{equation}
Similarly, we obtain
\begin{equation} \label{Nplus}
N_{-}|n\rangle = \sqrt{n+2}|n+2\rangle
\end{equation}
Thus, the operators $N_{-}$ and $N_{+}$ are generators of the square roots of natural numbers using Eqs. (\ref{Nminus}) and (\ref{Nplus}), starting from $\sqrt{2}$ the Pythagora's constant.

In this construction, we can successively apply the operator $N_{-}$ to the QNSV $|0\rangle$ and $|1\rangle$ to generate separately even and odd QNSV:
\begin{equation} \label{recursive}
\begin{split}
    |2\rangle  = \frac{N_{-}}{\sqrt{2}}|0\rangle, \quad |3\rangle  = \frac{N_{-}}{\sqrt{3}}|1\rangle \\
    |4\rangle  = \frac{(N_{-})^2}{\sqrt{2} \sqrt{4}}|0\rangle, \quad |5\rangle  = \frac{(N_{-})^2}{\sqrt{3} \sqrt{5}}|1\rangle \\
    |6\rangle  = \frac{(N_{-})^3}{\sqrt{2} \sqrt{4} \sqrt{6}}|0\rangle, \quad |7\rangle  = \frac{(N_{-})^3}{\sqrt{3} \sqrt{5} \sqrt{7}}|1\rangle
    \\
    \vdots\\
    |2n\rangle  = \frac{(N_{-})^n}{\sqrt{(2n)!!}}|0\rangle, \quad |2n+1\rangle  = \frac{(N_{-})^n}{\sqrt{(2n+1)!!}}|1\rangle
\end{split}
\end{equation}
for which we have introduced the definition of double factorials \cite{gould2012} in the leading coefficients.

With this partitioning, it is useful to separate the matrix representations of the operators $N_{+}$ and $N_{-}$ in the basis $\{ |n\rangle \}$ into different parities. For arbitrary finite representations, their even representations results, respectively, in matrices of the form 
\begin{equation} \label{evenrep}
    [N_{+}]^{e} = \begin{pmatrix}
         0 & \sqrt{2} & 0 & 0 & \dots & 0 \\
         0 & 0 & \sqrt{4} & 0 & \dots & 0 \\
         0 & 0 & 0 & \sqrt{6} & \dots & 0 \\
         0 & 0 & 0 & 0 & \ddots & \vdots\\
          \vdots & \vdots & \vdots & \vdots & \ddots & \sqrt{2n} \\
         0 & 0 & 0 & 0 & \dots & 0 \\
         
    \end{pmatrix}, \quad [N_{-}]^{e} = \begin{pmatrix}
         0 & 0 & 0 & 0 & \dots & 0 \\
         \sqrt{2} & 0 & 0 & 0& \dots & 0 \\
         0 & \sqrt{4} & 0 &0& \dots & 0\\ 
         0 & 0 & \sqrt{6} & 0&\dots & 0 \\
          \vdots & \vdots & \vdots &\ddots & \ddots & \vdots\\ 
         0 & 0 & 0 &\dots& \sqrt{2n}&0 \\
         
    \end{pmatrix}
\end{equation}
while their odd representations results, respectively, in matrices of the form
\begin{equation} \label{oddrep}
    [N_{+}]^{o} = \begin{pmatrix}
           0 & \sqrt{3} & 0 & 0 & \dots & 0 \\
         0 & 0 & \sqrt{5} & 0 & \dots & 0 \\
         0 & 0 & 0 & \sqrt{7} & \dots & 0 \\
         0 & 0 & 0 & 0 & \ddots & \vdots\\
          \vdots & \vdots & \vdots & \vdots & \ddots & \sqrt{2n+1} \\ 
         0 & 0 & 0 & 0 & \dots & 0 \\

    \end{pmatrix}, \quad [N_{-}]^{o} = \begin{pmatrix}
         0 & 0 & 0 & 0 & \dots & 0 \\
         \sqrt{3} & 0 & 0 & 0& \dots & 0 \\
         0 & \sqrt{5} & 0 &0& \dots & 0\\ 
         0 & 0 & \sqrt{7} & 0&\dots & 0 \\
          \vdots & \vdots & \vdots &\ddots & \ddots & \vdots\\
         0 & 0 & 0 &\dots& \sqrt{2n+1}&0 \\
    \end{pmatrix}
\end{equation}
As we have started the odd QNSV series from $|3\rangle$, note that both representations (even and odd) include zero and thus, using the definition of $N$, result 
\begin{equation} \label{e-o-rep}
    [N]^{e} = \begin{pmatrix}
         0 & 0 & 0 & \dots & 0 & \dots\\
         0 & 2 & 0 & \dots & 0 & \dots\\
         0 & 0 & 4 & \dots & 0 & \dots\\
          \vdots & \vdots & \vdots & \ddots & \vdots & \dots\\
         0 & 0 & 0 & \dots & 2n & \dots\\
         \vdots & \vdots & \vdots & \dots & \vdots & \ddots\\
    \end{pmatrix}, \quad [N]^{o} = \begin{pmatrix}
         0 & 0 & 0 & \dots & 0 & \dots\\
         0 & 3 & 0 & \dots & 0 & \dots\\
         0 & 0 & 5 & \dots & 0 & \dots\\
          \vdots & \vdots & \vdots & \ddots & \vdots & \dots\\
         0 & 0 & 0 & \dots & 2n+1 & \dots\\
         \vdots & \vdots & \vdots & \dots & \vdots & \ddots\\
    \end{pmatrix}
\end{equation}
In order to recover the faithful representation of $[N]$, we can write it as
\begin{equation} \label{totalNrep}
[N] = [N]^{e} \otimes \bigl[ \begin{smallmatrix}
  1&0\\ 0&0
\end{smallmatrix} \bigr] + \bigl[ \begin{smallmatrix}
  0&0\\ 0&1
\end{smallmatrix} \bigr] \otimes \left( [N]^{o} + \bigl[ \begin{smallmatrix}
  1&\textbf{0}\\ \textbf{0}&\textbf{0}
\end{smallmatrix} \bigr] \right)
\end{equation}
where $[N]^e$ and $[N]^o$ are matrices given by Eqs. (\ref{e-o-rep}), the matrices $\bigl[ \begin{smallmatrix}
  1&0\\ 0&0
\end{smallmatrix} \bigr]$ and $\bigl[ \begin{smallmatrix}
  0&0\\ 0&1
\end{smallmatrix} \bigr]$ are $2 \times 2$, and the matrix $\bigl[ \begin{smallmatrix}
  1&\textbf{0}\\ \textbf{0}&\textbf{0}
\end{smallmatrix} \bigr]$, containing only 1 in the first entry and 0 for all other elements, is used to regularize the odd representation.

\subsection{The concept of qu$n$it in the natural $q$-number basis}

Given an arbitrary ket $|Q \rangle$ in the space spanned by the QNSV of the natural $q$-number $N$, it is expanded as follows:
\begin{equation} \label{qnit2}
|Q\rangle = \sum_{n=0}^{\infty} c_{n}|n\rangle
\end{equation}
We denote this expansion as a qu$n$it, i.e., a `basic unit of number information' in the basis of the natural $q$-number, within QNT. In this case, we can define qu$n$its with $n \geq 2$, being a qu$2$it the simplest superposition in the natural basis. However, it is important to stress here that a qu$n$it is not a qudit, a concept currently used in quantum computing \cite{kiktenko2020}, although there is a close connection between these entities. 

By multiplying $\langle n'|$ on the left in both sides of Eq. (\ref{qnit2}), and using the orthonormality property $\langle n'|n\rangle = \delta_{n'n}$, we read an expansion coefficient as a projection of $|Q \rangle$ onto a specific QNSV, $c_{n'} = \langle n'|Q\rangle \in \mathbb{C}$.
If the superposition (\ref{qnit2}) is normalized, we define it as a qu$n$it of infinite dimension,
satisfying a convergent series $\sum_{n=0}^{\infty} |c_{n}|^2 = 1$. 

While QNT might be concerned with a complete Hilbert space of the natural $q$-number, it is convenient to approach it through $d$-dimensional subspaces:
\begin{equation}
\mathcal{H}^{d} = \left\{\sum_{n=0}^{d-1} \alpha_{n}|n\rangle; (\alpha_{n} \in \mathbb{C}) \wedge \left( \sum_{n=0}^{d-1} \big| \alpha_{n} \big|^{2} = 1 \right) \right\}
\end{equation}
At this point, we reinforce that a finite qu$n$it is not yet a qudit of quantum computing theory, which represents a $d$-dimensional quantum information unit \cite{erhard2018,wang2017,jafarzadeh2020}. We will try to do this connection in the following. We notice that within this QNT it is possible to define any number-theoretical superposition state, including for the primes. For instance, following Ref. \cite{sierra2020} we can write a prime qu$n$it (or a qu$p$it) in the form
\begin{equation} \label{qnitprime}
|Q'_{n}\rangle = \frac{1}{\sqrt{\pi(2^n)}} \sum^{2^n}_{p:primes} |p\rangle
\end{equation}
which corresponds to an equally likely quantum superposition of all prime numbers less than $2^n (n \geq 2)$, with $\pi(x)$ being the prime-number counting function that gives the number of primes smaller than or equal to $x$.

It is convenient first to discuss the concept of ``numbering'' or ``counting'' in QNT. We adopt a Dirac-type prescription \cite{dirac3}, where a qu$n$it represents a single QNSV superposition in which ``to number'' means to project onto a QNSV of a specific $q$-number (in this case, $N$), resulting in a specific natural $c$-number. In this sense, we interpret this situation as follows. Before a counting of the natural $q$-number is realized, a qu$n$it is assumed to be represented by some linear combination of the form (\ref{qnit2}). Thus, when a counting is performed, the qu$n$it is projected onto one of the QNSV, say $|n'\rangle$  of $q$-number $N$ resulting in $c$-number $n'$. However, we do not know in advance onto which of the several QNSV the qu$n$it will be projected as the result of this counting. Hence, we adopt the postulate that the quantum probability $P_{Q}(n')$ for projecting onto some specific $|n'\rangle$ obtaining $n'$ will be given by
\begin{equation} \label{Probability}
P_{Q}(n') = |\langle n'|Q\rangle|^{2}
\end{equation}
since that $|Q\rangle$ may be normalized.

Although it is not yet clear what does ``counting'' mean here, this type of numbering could be performed by a quantum processing \cite{reimer2019}. To determine the probability (\ref{Probability}) empirically, we could consider a great number of computations performed on a set of qu$n$its or ensemble. Let us introduce the concept of ``mixed qu$n$its'', a collection of kets $|Q^{(1)}\rangle$, $|Q^{(2)}\rangle$, $|Q^{(3)}\rangle$, ..., $|Q^{(k)}\rangle$, which need not be orthogonal. This is analogous to the notion of mixed states, originally introduced by von Neumann and Landau in quantum mechanics \cite{von-neumann}. In this sense, a completely random ensemble is a collection of equally likely qu$n$its. In principle, a ``pure qu$n$it'' $|Q\rangle$ can be represented by a one-dimensional subspace or ray in a Hilbert space. A completely random ensemble and a pure ensemble are considered as extremes of the cases known as mixed ensembles. 

In general, for the collection of $k$ distinct qu$n$its, we assume that a fraction of the elements with probability weight $w_{1}$ is characterized by $|Q^{(1)}\rangle$, some other fraction with probability weight $w_{2}$, by $|Q^{(2)}\rangle$, and so on. We also impose that the probability weights are constrained to satisfy the normalization condition $\sum_{i=1}^{k} w_{i} = 1$.
Thus, if a $w_{j} = 1$ for some $|Q^{(j)}\rangle$, while all others vanish, the set of qu$n$its represents a pure ensemble, where all elements exhibit the same superposition state. In general, the number of distinct qu$n$its needs not be coincident with the dimensionality $d$ of a natural QNSV subspace; i.e., $k \ge d$. We note that if $d = \infty$, the cardinality of the collection $\big\{|Q^{(1)}\rangle, |Q^{(2)}\rangle, |Q^{(3)}\rangle,..., |Q^{(k)}\rangle \big\}$ could be larger than the size of an infinite qu$n$it, such as it occurs in Cantor's paradox. However, we can declare that this collection is a proper class, in the sense of von Neumann-Bernays-G\"odel set theory \cite{von-neumann1928}. These concepts enable us to make a statistical interpretation of QNT. 

Let us suppose now that we can generalize the concept of numbering for a completely random ensemble of a $q$-number $N$. Then, we may ask what is the average numeric value of $N$ when a large number of computations are carried out? In this case, the answer is given by the ensemble average of $N$ in a subspace of dimension $d$, which is properly defined by
\begin{equation} \label{ave-ens}
\langle \langle N \rangle \rangle_{Q} \equiv \sum_{i=1}^{k} w_{i} \langle Q^{(i)}|N|Q^{(i)}\rangle = \sum_{i=1}^{k} \sum_{n'=0}^{d-1} w_{i} |\langle n'|Q^{(i)}\rangle|^2 \hspace{0.1cm} n'
\end{equation}
where $|n'\rangle$ is any $d$-dimensional QNSV of $N$ associated to the $c$-number $n'$. Here, it is important to highlight that in Eq. (\ref{ave-ens}), $|c_{n'}|^2 =|\langle n'|Q^{(i)}\rangle|^2 = P_{Q}(n')$ can be understood as a quantum probability (``$q$-probability'') for a qu$n$it $|Q^{(i)}\rangle$ to be projected onto QNSV $|n'\rangle$, whereas the real $c$-number $w_{i}$ can be understood as a classical probability factor (a ``$c$-probability'') of finding in the ensemble a qu$n$it $|Q^{(i)}\rangle$. 

We are able now to introduce the von Neumann density operator formalism in QNT. Thus, we define the Hermitian number density operator $\rho_{Q}$ as follows
\begin{equation} \label{gendmatrix}
\rho_{Q} = \sum_{i=1}^{k} w_{i} |Q^{(i)}\rangle \langle Q^{(i)}|
\end{equation}
The elements of the corresponding number density matrix in the natural basis have the form 
\begin{equation} \label{eledmatrix}
\langle n''|\rho_{Q}| n' \rangle = \rho_{n''n'} = \sum_{i=1}^{k} w_{i} \langle n''|Q^{(i)}\rangle \langle Q^{(i)}|n' \rangle
\end{equation}
such that $\text{Tr}(\rho) = \sum_{n'}^{} \rho_{n'n'}$ is 
\begin{equation*} \label{trace}
\sum_{i=1}^{k} \sum_{n'=0}^{d-1} w_{i} \langle n'|Q^{(i)}\rangle \langle Q^{(i)}|n' \rangle = \sum_{i=1}^{k} w_{i} \langle Q^{(i)}|Q^{(i)}\rangle = 1
\end{equation*}
Note that, for a pure ensemble $\rho_{Q} = |Q\rangle \langle Q|$, $\rho_{Q}(\rho_{Q} - I) = 0$, and $\text{Tr}(\rho^{2}) = 1$. The diagonalized number density matrix for a pure $d$-dimensional qu$n$it is represented by a $d \times d$ matrix of the form
\begin{equation} \label{purematrix}
    [\rho_Q]_{pure} = \left( \begin{array}{cccccc}
         0 & & & 0 & \dots & 0\\
         0 & & & \ddots &  &0\\
         \vdots &  & & & 1 &\vdots\\
         0 & & & 0 & \dots & 0\\
    \end{array}\right)
\end{equation}
Therefore, each pure qu$n$it can be represented by a matrix that in diagonal form consists of zeros everywhere except for a single 1 on the diagonal. To find a mixture of qu$n$it states, a set of $d \times d$ matrices exhibiting different nonzero diagonal is required. Thus, the density matrix for a completely random ensemble is given by
\begin{equation} \label{randommatrix}
    [\rho_Q]_{random} = \frac{1}{d} \left( \begin{array}{cccccc}
         1 & & & 0 & \dots & 0\\
         0 & & & \ddots &  &0\\
         \vdots &  & & & 1 &\vdots\\
         0 & & & 0 & \dots & 1\\
    \end{array}\right)
\end{equation}

After that, we can discuss on the connection between the density operator formalism and the statistical QNT. Following the quantum-mechanical formalism, to quantify the difference of numeric information between a completely random ensemble and a pure ensemble, we use the quantity $\Omega$ defined, within a basis in which $\rho$ is diagonal, by
\begin{equation} \label{information}
\Omega = -\sum_{k}^{} \rho^{diag}_{kk}\ln{\rho^{diag}_{kk}}
\end{equation}
As in quantum mechanics, each element $\rho^{diag}_{kk}$ is a real number between 0 and 1, such that $\Omega$ is necessarily positive semidefinite ($\Omega \ge 0$). Hence, for a completely random ensemble of qu$n$its, it gives 
\begin{equation} \label{random}
\Omega = -\sum_{k=1}^{d} \frac{1}{d} \ln{\frac{1}{d}} = \ln{d}
\end{equation}
Otherwise, for a pure ensemble we read $\Omega = 0$, where we have utilized $\rho^{diag}_{kk} = 0$ or $\ln{\rho^{diag}_{kk}} = 0$. 

In this statistical interpretation of QNT, we can argue that $\Omega$ can represent a quantitative measure of randomness of the ensemble. Thus, a pure ensemble of qu$n$its is a class where all elements are characterized by the same superposition defined as $|Q \rangle$ in Eq. (\ref{qnit2}) with $\Omega = 0$. On the other hand, a completely random ensemble, in which all elements are equally likely, exhibits large value for $\Omega$ when $d \shortrightarrow \infty$, being maximal when $\sum_{n}^{} \rho_{nn} = 1$.

Let us consider in more detail an ensemble of $q$-number $N$. Let $N$ be a collection of independent natural $q$-numbers $N^{(1)}$, $N^{(2)}$, $N^{(3)}$, ..., $N^{(k)}$, with $[N^{(i)},N^{(j)}]=0$, for all pairs $i,j$. In this case, the number density operator can be generalized as
\begin{equation} \label{ggendmatrix}
\rho_{Q} = \sum_{n'_{1} ... n'_{k}}^{} |n'_{1}...n'_{k}\rangle w_{n'_{1} ... n'_{k}} \langle n'_{1}...n'_{k}|
\end{equation}
where $\big\{|n'_{i}\rangle \big\}$ is a basis of QNSV belonging to $N^{(i)}$ ($i=1,2,...,k)$. Here, we have respected the statistical independence, i.e., we assume that the natural $q$-numbers are distinguishable. However, if the $q$-numbers are indistinguishable, we need to improve the formulation by imposing symmetry considerations. We will not treat this latter case here. Note that in Eq. (\ref{ggendmatrix}), we can write
\begin{equation}
w_{n'_{1} ... n'_{k}} = w_{1}(n'_{1}) w_{2}(n'_{2})... w_{k}(n'_{k})
\end{equation}
since we can define the QNSV product
\begin{equation}
|n'_{1}...n'_{k}\rangle =  |n'_{1}\rangle |n'_{2}\rangle...|n'_{k}\rangle
\end{equation}
Thus, the number density operator is given by
\begin{equation} 
\rho_{Q} = \prod_{i=1}^{k} \rho_{i}, \quad \rho_{i} = \sum_{n'_{i}}^{} |n'_{i}\rangle w_{i} \langle n'_{i}|
\end{equation}
With this new definition of uniform ensemble of natural $q$-numbers, we recover Eqs. (\ref{information}) and (\ref{random}) that enable us to speak about the concept of `quantum number information'.

For a distinguishable and equally probable ensemble, it is also convenient to measure the quantity of information in classical bits. Hence, using Eq. (\ref{random}), we introduce the notion of Shannon information quantity \cite{nielsen2010}, $I = K\ln{d}$, for which we set $K=1/\ln{2}$, such that
\begin{equation} \label{shannon}
I_{clas.} = \log_{2}{d}
\end{equation}
where $d$ is the dimensionality of the QNSV space.

\subsection{Quantifying the $q$-probability distribution for a qu$n$it state}

Let us consider again an arbitrary qu$n$it state $|Q\rangle$ of the form (\ref{qnit2}). In order to determine the expansion coefficients and, consequently, the $q$-probability for projecting onto some specific natural QNSV, we consider that a qu$n$it is associated with a complex number of the form $q = n_1 + in_2$. As we examine below, $n_1$ and $n_2$ do not necessarily have to be integers (cf. Eq. (\ref{nplus})). Thus, we look for a qu$n$it that is an eigenvector of operator $N_{+}$, i.e.,
\begin{equation} \label{eigenqunit}
N_{+}|Q\rangle = q|Q\rangle
\end{equation}
where $q \in \mathbb{C}$. In this sense, $q$ is a complex $c$-number belonging to $N_{+}$, which should be in some extent a complex $q$-number.

Now, we can determine the coefficients $c_{n}$ of Eq. (\ref{qnit2}) by applying $N_{+}$, remembering Eq. (\ref{Nminus}),
\begin{equation} 
N_{+}|Q\rangle = \sum_{n=0}^{\infty} c_{n}N_{+}|n\rangle = \sum_{n=2}^{\infty} c_{n}\sqrt{n}|n-2\rangle
\end{equation}
After the shift $n-2 \xrightarrow{} n$ of summation variable, this becomes
\begin{equation}
N_{+}|Q\rangle = \sum_{n=0}^{\infty} c_{n+2}\sqrt{n+2}|n\rangle
\end{equation}
Comparing with the original expansion for a qu$n$it, we obtain 
\begin{equation*} \label{cnrelation}
c_{n+2} = \frac{q}{\sqrt{n+2}}c_{n}
\end{equation*}
This latter relation enables us to determine by recurrence all the coefficient separately in terms of $c_{0}$ and $c_{1}$, i.e.,
\begin{equation} \label{recursion}
c_{2n} = \frac{q^{n}}{\sqrt{(2n)!!}}c_{0}, \quad c_{2n+1} = \frac{q^{n}}{\sqrt{(2n+1)!!}}c_{1}
\end{equation}
It follows that, when $c_{0}$ is fixed, all even $c_{n}$ are also fixed and when $c_{1}$ is fixed, all odd $c_{n}$ are also fixed. 

In this basis, the expansion of a qu$n$it can be separated into parity even and odd
\begin{equation} \label{Qnit}
|Q\rangle = c_{0}\sum_{n=0}^{\infty} \frac{q^{n}}{\sqrt{(2n)!!}}|2n\rangle + c_{1}\sum_{n=0}^{\infty} \frac{q^{n}}{\sqrt{(2n+1)!!}}|2n+1\rangle
\end{equation}
i.e., with the partitioning $|Q\rangle = |Q_{2k}\rangle + |Q_{2k+1}\rangle$. Equivalently, it implies the existence of a decomposition of the Hilbert space $\mathcal{H}$ of the qu$n$its into a direct sum of subspaces with parity even and odd, i.e., $\mathcal{H} = \mathcal{H}_{2k} \oplus \mathcal{H}_{2k+1}$.

Note that if we choose $c_{0}$ real and positive and $c_{1}=0$, Eq.(\ref{Qnit}) reduces to a projection onto a subspace spanned by even QNSV, including the ket $|0\rangle$:
\begin{equation} \label{Qodd}
|Q_{2k}\rangle = c_{0}\sum_{n=0}^{\infty} \frac{q^{n}}{\sqrt{(2n)!!}}|2n\rangle
\end{equation}
Normalizing the ket $|Q_{2k}\rangle$, we obtain (using $(2n)!! = 2^{n}n!$ for $n=0,1,\dots$)
\begin{equation} \label{Qodd1}
|c_{0}|^{2}\sum_{n=0}^{\infty} \frac{|q|^{2n}}{(2n)!!} = |c_{0}|^{2}\sum_{n=0}^{\infty} \frac{1}{n!} \left( \frac{|q|^{2}}{2} \right)^{n} = |c_{0}|^{2}e^{|q|^2/2} = 1
\end{equation}
With this convention, we have chosen:
\begin{equation*} \label{c0fixed}
c_{0} = e^{-|q|^2/4}
\end{equation*}
and, therefore
\begin{equation} \label{Qodd2}
|Q_{2k}\rangle = e^{-|q|^2/4}\sum_{n=0}^{\infty} \frac{q^{n}}{\sqrt{(2n)!!}}|2n\rangle
\end{equation}

We notice, from Eq. (\ref{Qodd2}), that the $q$-probability of projecting $|Q_{2k}\rangle$ onto an arbitrary even QNSV, resulting in an even $c$-number, is given by
\begin{equation} \label{evenpoisson}
P_{even}(q) = |c_{2n}|^2 = \frac{1}{n!} \left( \frac{|q|^{2}}{2} \right)^{n} e^{-|q|^2/2}
\end{equation}
By choosing $\lambda = |q|^2/2$, we recognize Eq. (\ref{evenpoisson}) as a Poisson distribution, which gives which gives a statistically independent distribution of even numbers, including zero. Although a similar distribution is well known for coherent states in quantum mechanics, in QNT it may be related with the notion of probability distribution on the natural numbers \cite{kerkvliet2016}, in a context of a probabilistic QNT.

The Poisson distribution reaches its maximum value when $n$ equal the integral part of $|q|^2/2$; i.e., it obeys a Pythagorean relation in a complex circle. For example, we can calculate the mean value of the natural $q$-number $N^{\star}$ in the even sector by using Eq. (\ref{evenpoisson}) in the expression:
\begin{equation}
\langle \hspace{0.0cm} N^{\star} \hspace{0.0cm}  \rangle_{Q_{2k}} = \sum_{n=0}^{\infty} P_{even}(q) \left( n+1 \right)
\end{equation}
Notwithstanding, it is simpler to use the adjoint relation of Eq. (\ref{eigenqunit}):
\begin{equation} \label{eigenadjoint}
\langle Q|N_{-} = q^{*}\langle Q|
\end{equation}
in such way that
\begin{equation}
\langle Q_{2k}|N_{-}N_{+}|Q_{2k} \rangle = \langle Q_{2k}|N|Q_{2k} \rangle = |q|^2 
\end{equation}
and therefore the even mean value of $N^{\star}$ is
\begin{equation}
\langle {N^{\star}} \rangle_{Q_{2k}} = \langle Q_{2k}|N + I|Q_{2k} \rangle = n_1^{2} + n_2^{2} + 1
\end{equation}
for which several pairs of $n_1$ and $n_2$ are possible as solutions in the complex plane. In particular, the unit circle ($n_1^{2} + n_2^{2} = 1$) is a Pythagorean solution for $\langle {N^{\star}} \rangle_{Q_{2k}} = 2$.

Otherwise, if we consider a qu$n$it spanned by odd QNSV, i.e.,
\begin{equation} \label{Qeven}
|Q_{2k+1}\rangle = c_{1}\sum_{n=0}^{\infty} \frac{q^{n}}{\sqrt{(2n+1)!!}}|2n+1\rangle
\end{equation}
we obtain, by normalizing,
\begin{equation} \label{Qeven1}
|c_{1}|^{2}\sum_{n=0}^{\infty} \frac{|q|^{2n}}{(2n+1)!!} = \frac{|c_{1}|^{2}}{|q|}\sum_{n=0}^{\infty} \frac{|q|^{2n+1}}{(2n+1)!!} = 1
\end{equation}
For this case, we note from Ref. \cite{gould2012} that
\begin{equation}
\sum_{n=0}^{\infty} \frac{x^{2n+1}}{(2n+1)!!} = \sqrt{\frac{\pi}{2}} \text{erf} \left(\frac{x}{\sqrt{2}}\right)e^{x^{2}/2}
\end{equation}
Thus, the coefficient $c_{1}$ can be calculated from 
\begin{equation*}
\frac{|c_{1}|^{2}}{|q|}\sqrt{\frac{\pi}{2}} \text{erf} \left(\frac{|q|}{\sqrt{2}}\right)e^{|q|^{2}/2} = 1
\end{equation*}
such that
\begin{equation*}
c_{1} = f(|q|) e^{-|q|^{2}/4}
\end{equation*}
where $f(|q|) = |q|^{1/2}/\left(\sqrt{\frac{\pi}{2}} \text{erf} \left(\frac{|q|}{\sqrt{2}}\right)\right)^{1/2}$, and
\begin{equation}
|Q_{2k+1}\rangle = f(|q|) e^{-|q|^{2}/4}\sum_{n=0}^{\infty} \frac{q^{n}}{\sqrt{(2n+1)!!}}|2n+1\rangle
\end{equation}
The $q$-probability distribution of projecting $|Q_{2k+1}\rangle$ onto an arbitrary odd QNSV in this case reads (observing that $(2n+1)!! = (2n+1)!/2^{n} n!$ for $n=0,1,...$)
\begin{equation} \label{oddpoisson}
P_{odd}(q) = |c_{2n+1}|^2 = \frac{|q|^{2n+1}}{(2n+1)!} \frac{2^{n}n!\hspace{0.1cm}e^{-|q|^2/2}}{\sqrt{\frac{\pi}{2}} \text{erf} \left(\frac{|q|}{\sqrt{2}}\right)}
\end{equation}
After a convenient manipulation of Eq. (\ref{oddpoisson}), we find
\begin{equation} \label{oddpoisson2}
    P_{odd}(q)=\frac{|q|2^{2n}(n!)^{2}}{(2n+1)!\sqrt{\frac{\pi}{2}}\text{erf}\left(\frac{|q|}{\sqrt{2}}\right)}\left[\frac{1}{n!}\left(\frac{|q|^{2}}{2}\right)^{n}e^{-|q|^{2}/2}\right]
\end{equation}
for which the expression inside the square brackets is recognized as a Poisson distribution, by taking $\lambda = |q|^{2}/2$ exactly as in Eq. (\ref{evenpoisson}).

\begin{figure}[ht]
	\centering
	\includegraphics[width=8cm,height=11cm]{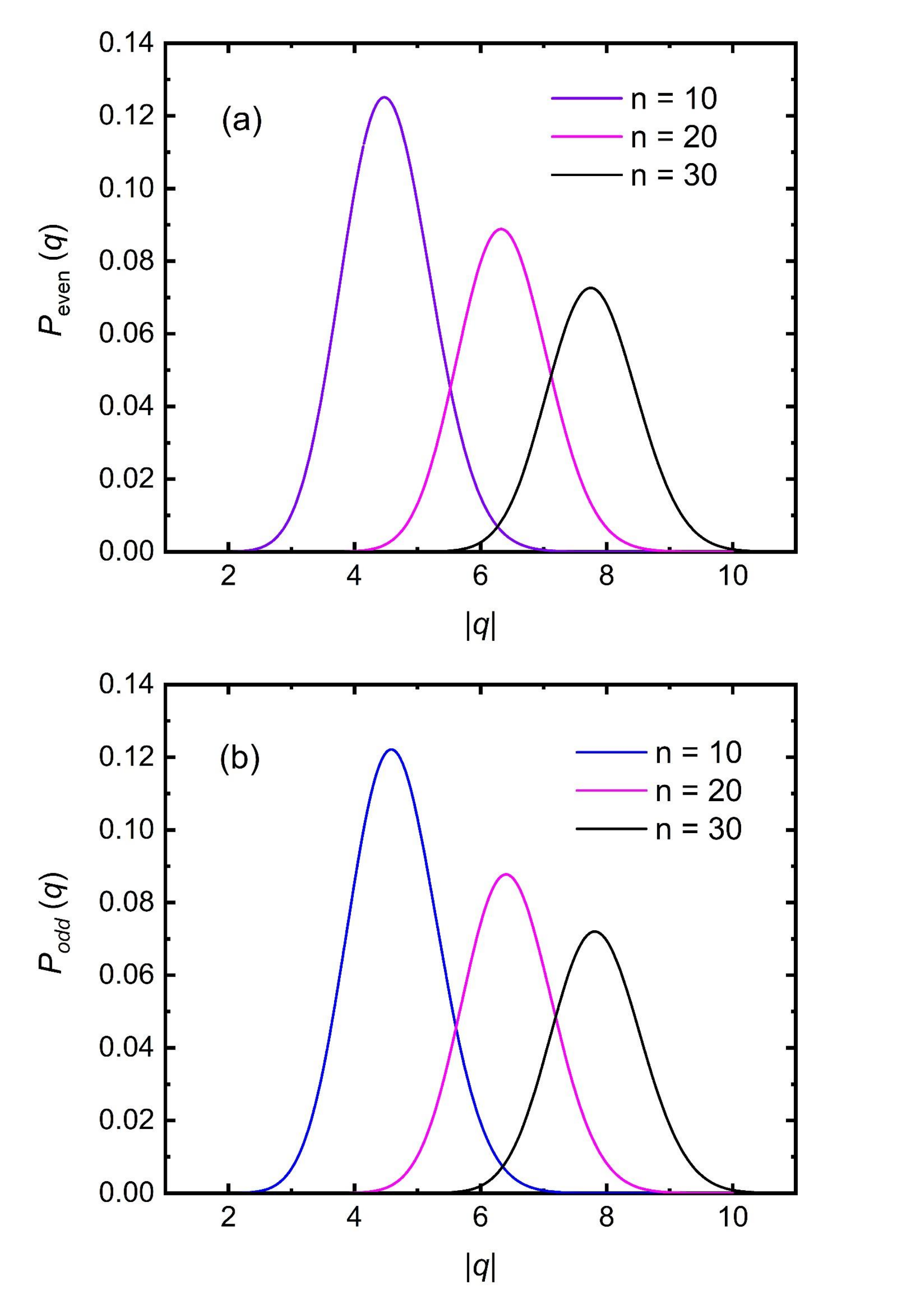}
	\caption{Calculated (a) even and (b) odd number probability distributions using Eqs. (\ref{evenpoisson}) and (\ref{oddpoisson2}), respectively.}
	\label{pdist}
\end{figure}

It is interesting to relate both distributions. Hence, using the Stirling's approximation, $n!\sim\sqrt{2\pi n}\left(\frac{n}{e}\right)^{n}$, we can write an approximate relationship between the even and odd distributions
\begin{equation}
    P_{odd}(q) \approx \frac{|q|A_{n}}{\sqrt{\frac{\pi}{2}}\text{erf}\left(\frac{|q|}{\sqrt{2}}\right)}P_{even}(q)
\end{equation}
with 
\begin{equation*}
    A_{n} = \frac{\pi(2n)^{2n+1}}{(2n+1)!} e^{-2n}
\end{equation*}
or, in a more general way, 
\begin{equation}
    P_{odd}(q) \approx \frac{|q|B_n}{\text{erf}\left(|q|/\sqrt{2}\right)} P_{even}(q)
\end{equation}
with
\begin{equation*}
    B_n = \frac{e}{\sqrt{2n+1}}\left(\frac{2n}{2n+1}\right)^{2n+1}
\end{equation*}

From this analysis, the probability distribution of projecting a qu$n$it of $\mathcal{H}_{2k+1}$ onto an odd QNSV does not appear exactly a Poisson distribution, as obtained for the probability distribution of projecting a qu$n$it of $\mathcal{H}_{2k}$ onto an even QNSV, including $|0\rangle$. Indeed, this result is not simple to interpret classically, because there is a bijection between both $\mathbb{N}$ and $2\mathbb{N}$ and $\mathbb{N}$ and $\mathbb{Z}_{+} = \big\{1,2,3,...\big\}$; i.e., these sets have the same cardinality. Notwithstanding, if we plot both distributions $P_{even}$ and $P_{odd}$ (cf. Figure 1), for different values of $n$, we notice that they represent the same class of probability distribution, as is expected for all natural numbers \cite{schirokauer2007}, converging for large values of $n$ and giving equal probabilities to each natural $c$-number, regardless of its parity.

\subsection{The simplest qu$n$it in the natural $q$-number basis}

As a simple application of the concept of qu$n$it, we write now a qu2it as a superposition in a 2D Hilbert space:
\begin{equation} \label{Q2it} 
|Q_{2}\rangle = c_{0}|0\rangle + c_{1}|1\rangle
\end{equation}
satisfying the property $c^2_{0} + c^2_{1} = 1$. Note that this qu$n$it is now a superposition of two QNSV of $N$ with parity even and odd, respectively. Here, we can choose $c_{0}$ and $c_{1}$ as being arbitrary complex $c$-numbers. Thus, we rewrite Eq. (\ref{Q2it}) in the polar form
\begin{equation} \label{Qcomplex}
|Q_{2}\rangle = e^{i\gamma}[\cos(\theta/2) |0\rangle +  e^{i\phi} \sin(\theta/2)|1\rangle]
\end{equation}
where $0 \leq \gamma < 2\pi$, $0 \leq \phi < 2\pi$, and $0 \leq \theta \leq \pi$.
Note that the QNSV basis $|0\rangle$ and $|1\rangle$ are eigenvectors of the natural $q$-number $N$
(${N}^{\star}$) associated to the natural $c$-numbers 0 and 1 (1 and 2), respectively. In quantum computing, the $c$-number $e^{i\gamma}$ in Eq. (\ref{Qcomplex}) is denoted a global phase factor and $|Q_{2}\rangle$ is denoted a qubit \cite{nielsen2010}.

It is useful here to use the geometric representation for a usual qubit in $\mathbb{R}^3$. In this sense, qu$2$its of the form $|Q_{2}\rangle_{R} = \cos(\theta/2) |0\rangle +  e^{i\phi} \sin(\theta/2)|1\rangle$ belong to a three-dimensional (3D) vector subspace $\mathbb{C}^{2}(\mathbb{R})$. Now, it can be rewritten as
\begin{equation} \label{Qreal}
|Q_{2}\rangle_{R} = \cos(\theta/2) \left( \begin{array}{cc}
         1\\
         0\\
    \end{array}\right) +  \cos\phi \sin(\theta/2) \left( \begin{array}{cc}
         0\\
         1\\
    \end{array}\right) + \sin\phi \sin(\theta/2) \left( \begin{array}{cc}
         0\\
         i\\
    \end{array}\right)
\end{equation}
for which we have chosen the basis
\begin{equation*}
\Bigg\{\left( \begin{array}{cc}
         1\\
         0\\
    \end{array}\right), \left( \begin{array}{cc}
         0\\
         1\\
    \end{array}\right), \left( \begin{array}{cc}
         0\\
         i\\
    \end{array}\right) \Bigg\}
\end{equation*}
Since this subspace is defined on the real field, it is isomorphic to $\mathbb{R}^3$, such that the transformation $T(|Q_{2}\rangle_{R})=|Q_{3D}\rangle_{R}$ leads to a 3D representation,
\begin{equation} \label{Qreal3D}
|Q_{3D}\rangle_{R} = \cos(\theta/2) \left( \begin{array}{cc}
         1\\
         0\\
         0\\
    \end{array}\right) +  \cos\phi \sin(\theta/2) \left( \begin{array}{cc}
         0\\
         1\\
         0\\
    \end{array}\right) + \sin\phi \sin(\theta/2) \left( \begin{array}{cc}
         0\\
         0\\
         1\\
    \end{array}\right)
\end{equation}
Note that $|Q_{3D}\rangle_{R}$ is an isomorphism and not a qutrit \cite{kurzynski2016,scirep2013}. This new vector represents a hemisphere in $\mathbb{R}^3$, denoted $SE^{2}$, of unit radius and centered at the origin. Hence, we can map a superposition of two natural QNSV, $|0\rangle$ and $|1\rangle$, into $\mathbb{R}^3$.  

A well-known way to represent a vector such as $|Q_{2}\rangle$ is using the Bloch sphere representation. In this case, the Bloch vectors span the whole unit ball in $\mathbb{R}^3$, for which any point $\boldsymbol{r}=(x_1,x_2,x_3)$ with $|\boldsymbol{r}| \leq 1$ corresponds to an arbitrary qu2it. Only points on the surface ($|\boldsymbol{r}| = 1$) are of the form defined by Eq. (\ref{Q2it}), i.e., pure qu2it states, whereas points inside the ball ($|\boldsymbol{r}| < 1$) correspond to qu2it mixed states. Thus, using Eq. (\ref{Qcomplex}) and the Hopf coordinate transformation, we define the number density matrix $\rho_{Q}$ of a pure qu$2$it state given by $\rho_{Q} = |Q_{2}\rangle \langle Q_{2}|$ (cf. Eq. (\ref{gendmatrix})). More specifically, we obtain
\begin{equation} \label{densitymatrix}
\rho_{Q} = \frac{1}{2} \left( \begin{array}{cccccc}
         1+x_{3} & x_{1}-ix_{2} \\
         x_{1}+ix_{2} & 1-x_{3} \\
    \end{array}\right) = \frac{1}{2} \left(I + x_{1}\sigma_{1} + x_{2}\sigma_{2} + x_{3}\sigma_{3} \right)
\end{equation}
where we have introduced the Pauli vector $\boldsymbol{\sigma} = (\sigma_{1}, \sigma_{2}, \sigma_{3})$, such that in the compact version we have $\rho_{Q} = \frac{1}{2}(I + \boldsymbol{r} \cdot \boldsymbol{\sigma})$.

To end this Section, we notice that the natural $q$-number description proposed in this QNT allows to obtain a proper orthonormal basis in terms of natural QNSV that is separated by parity. This is, a general qu$n$it can be decomposed as $|Q\rangle = |Q_{2k}\rangle + |Q_{2k+1}\rangle$. By similarity, QNT is related to quantum computing theory and enables computations in higher dimensions than the 2D qubits and 3D qutrits. However, the construction of quantum logic gates operating on multiple qu$n$its is not a direct task in QNT. This is simpler in the case of a qu2it, which is an analogue of a qubit. Indeed, to treat entanglement and teleportation as a multidimensional case \cite{llewellyn2020}, we have to be careful with this algebraic construction. In the following, we show a more feasible way to deal with higher-order qu$n$its and other quantum operators. 

\section{The Integer Number Representation}

For completeness of QNT, we employ a Lie algebra case to define an integer $q$-number $\mathbf{Z}$, useful to generate $\mathbb{Z} = \big\{...,-3,-2,-1,0,+1,+2,+3,...\big\}$ and $\frac{1}{2}\mathbb{Z}^{*}=\big\{..., -3/2, -1/2, +1/2, +3/2, ...\big\}$, by exploiting this algebraic structure. We assume that the $Z$-space is isotropic, such that a $q$-number degeneracy naturally emerges from this structure. Thus, the meaning of the QNSV associated to \textbf{Z} is further analyzed via a proposed quantum mapping (cf. Section IV). Furthermore, a connection between the $Z$-representation and the SU($n$) group is performed to discuss our results in terms of higher-dimension quantum computing (cf. Section V).

\subsection{A 3D construction of the integer $q$-number}

\textbf{Axiom 2.}
Every classical integer number (namely, an integer $c$-number) is an eigenvalue of an integer quantum number operator (namely, an integer $q$-number).

\textbf{Definition 3.}
Let $\mathbf{Z} \equiv (Z_{1}, Z_{2}, Z_{3})$ be a 3D representation of the integer $q$-number in a Hilbert space satisfying the properties:
\begin{equation} \label{property4}
    \mathbf{Z}^{2} \equiv Z_{1}^{2} + Z_{2}^{2} + Z_{3}^{2}
\end{equation}
\begin{equation} \label{property5}
    \mathbf{Z} = \mathbf{Z}^{\dagger}
\end{equation}
\begin{equation} \label{property6}
    [Z_{i}, Z_{j}] = Z_{i}Z_{j} - Z_{j}Z_{i} = i\epsilon_{ijk}Z_{k}
\end{equation}
\begin{equation} \label{property7}
    [\mathbf{Z}^2, Z_{i}] = 0, \quad (i,j,k=1,2,3)
\end{equation}
Note that this definition is also valid for a half-integer quantum number representation, which generates $\frac{1}{2}\mathbb{Z}^{*}$ and has, as a particular case, the usual Pauli algebra, e.g., taking $\mathbf{Z}=(1/2)\boldsymbol{\sigma}$. Property (\ref{property4}) indicates that $\mathbf{Z}$ has a 3D representation in a Hilbert space; property (\ref{property5}) is chosen to obtain a Hermitian representations of the $q$-numbers; property (\ref{property6}) is to preserve the Lie algebra structure; and property (\ref{property7}) is to confirm that $\mathbf{Z}^2$ is also a Casimir operator, which can be generalized in the SU($n$) group.

From the above definition of $\mathbf{Z}$, we choose $Z_{3}$ as a preferential direction in a discrete Euclidean space. Thus, we conveniently write the the basic eigenvalue equations of QNT as
\begin{equation} \label{ztwo}
\mathbf{Z}^2|n,m\rangle = n(n+1)|n,m\rangle
\end{equation}
\begin{equation} \label{zk}
Z_{3}|n,m\rangle = m|n,m\rangle
\end{equation}
For these eigenvalue problems, the $n(n+1)$ are associated to the eigenvalues of $\mathbf{Z}^2$ and $m$ to the eigenvalues of $Z_{3}$, being $n$ a classical natural number ($n = 0, 1, 2, 3, ...$) and $m$ a classical integer ($m = 0, \pm 1, \pm 2, \dots, \pm n$). We notice again that this construction also allows half-integer solutions ($n = 1/2, 3/2, 5/2,...$) if the dimension of the configuration subspace is even. However, QNT should primarily deal with integer $q$-numbers and $c$-numbers, which could lead to a totally symmetric construction of this theory. Here, we also interpret the common kets $|n,m\rangle$ in Eqs. (\ref{ztwo}) and (\ref{zk}) as QNSV, which define orthonormal basis sets and span Hilbert spaces of $d$-dimensions, $d=2n+1$.

From the $\mathbf{Z}$ components, it is convenient to introduce the following non-Hermitian ladder operators
\begin{equation} \label{zplus}
    Z_{+} \equiv Z_{1} + iZ_{2}
\end{equation}
\begin{equation} \label{zminus}
    Z_{-} \equiv Z_{1} - iZ_{2}
\end{equation}
such that $[Z_{+},Z_{-}] = 2Z_{3}$ and $[Z_{-},Z_{+}] = -2Z_{3}$.
Therefore, $Z_{+}$ and $Z_{-}$ have the property: if $|n,m\rangle$ is a QNSV associated to the integer $c$-number $m$ of $Z_{3}$ and $|m|<n$, then 
\begin{equation}
\begin{array}{cc}
     Z_{+}|n,m\rangle = c_{nm}^{+}|n,m+1\rangle \\
     Z_{-}|n,m\rangle = c_{nm}^{-}|n,m-1\rangle \\
\end{array}
\end{equation}
In fact, the maximum (minimum) value of $m$, i.e., $\overline{m}$ ($\underline{m}$), is such that $Z_{+}|n,\overline{m}\rangle = 0$ ($Z_{-}|n,\underline{m}\rangle = 0$). Incidentally, these relations lead to
\begin{equation}
\begin{array}{cc}
     |c_{nm}^{+}|^2 = (n-m)(n+m+1) \\
    |c_{nm}^{-}|^2 = (n+m)(n-m+1) \\
\end{array}
\end{equation}

Using this basis set of QNSV, it is possible to define distinct units of quantum information that can be realized by a quantum system in terms of a superposition of mutually orthogonal quantum states \cite{dogra2018,nisbet2013}. These arbitrary vectors will be also called qu$n$its, with $n=0,1,2,...$, instead of using the nomenclature qudits for a $d$-dimensional space (commonly employed in quantum information \cite{liu2017}). Note that in the $Z$-representation, the dimension of the configuration space is fixed by $n$. Of course, there is also a connection between the qu$n$its associated to $q$-numbers $N$ (and $N^{\star}$) and $\mathbf{Z}$, by using a 3D representation for $\mathbf{N}$ inspired in the Schwinger algebra \cite{kumar2015}. However, in order to avoid a lengthy digression, we will not make such a connection here. 

In the $Z$-representation, our numeric description also includes a qu0it that defines a one-dimensional space, although it does not contain any quantum information. The next qu$n$it in this basis is the qu1it that is a superposition of three mutually orthonormal QNSV fixed by the natural $c$-number 1, which has been physically realized as a qutrit \cite{luo2019}. Thus, we obtain all higher dimensions qu$(2n+1)$its within this representation.

As is known for a true qutrit, every transition between pairs of the three state-vectors, $|1,-1\rangle$, $|1,0\rangle$ and $|1,+1\rangle$, should be independent. Hence, not all three-level quantum systems can be considered true qutrits. In the numeric description, if we restrict $\mathbf{Z}$ in such way that the matrix elements of $Z_{-}$ are  $\langle 1,-1 |Z_{-}| 1,0 \rangle = \langle 1,0 |Z_{-}| 1,1 \rangle = \sqrt{2}$, as in the case of the angular momentum operator, the transition between adjacent QNSV are not independent. In this sense, a qu1it can be thought as a pseudo-qutrit. (See, e.g., Refs. \cite{vaziri2003,fickler2014} for a proper definition of a true qutrit). In Section V, we discuss a possible connection between a qu1it and and a qutrit in more details.

It is also noticed that, in this proposed integer $q$-number basis, a qubit is a unit of quantum information belonging to the half-integer representation with $n=1/2$, i.e., a qu$\frac{1}{2}$it. In this case, we obtain an adequate description for a quantum computing based on qubits, since the QNSV are eigenvectors of $\mathbf{Z}=(1/2)\boldsymbol{\sigma}$, which constitute quantum logic Pauli gates \cite{nielsen2010}. More importantly, in this construction there is a possibility of extending the even-representation for higher dimensions, and defining quantum superpositions such as qutetrits ($n=3/2$), quhexits ($n=5/2$), etc. In this way, QNT can be connected with quantum computing theory related to higher dimensions than the usual quantum units. 

Now, we focus on the integer description, for which we have qu$n$its of the form
\begin{equation} \label{qnit}
|Q_{n}\rangle = \sum_{m=-n}^{+n} C_{nm}|n,m\rangle, \quad (n = 0, 1, 2, ...) 
\end{equation}
with $\sum_{m=-n}^{+n} \big| C_{nm} \big|^2 = 1$. This is known as the most general quantum pure state in the Majorana representation \cite{majorana} (see, e.g., Ref. \cite{dogra2018}). Notice that in quantum information, Weyl operators \cite{bertlmann2008} can be used as a type of Pauli basis for these qu$n$its, which can be properly constructed from the $\mathbf{Z}$ components (cf. Section V). 

\subsection{The algebraic numbers associated to the $\mathbf{Z}$ components}
To obtain a matrix representation of the integer $q$-numbers, we choose a finite canonical basis for the QNSV associated to the $q$-number $Z_{3}$. Thus, the column vectors belonging to the vector field $\mathds{V}$ over $\mathbb{C}$ span a $(2n+1)$-dimensional Hilbert space
\begin{equation} \label{canonical}
    \mathds{V} = \left[ \left( \begin{array}{cc}
         1\\
         0\\
         \vdots\\
         0\\
         
    \end{array}\right), \left( \begin{array}{cc}
         0\\
         1\\
         \vdots\\
         0\\
    \end{array}\right), \dots , \left(\begin{array}{cc}
         0\\
         0\\
         \vdots\\
         1\\
    \end{array}\right)  \right]
\end{equation}
As is usual, the row vectors can be also defined in the dual vector space since we need to define an inner product. By convention, we choose $Z_{3}$ to exhibit eigenvectors of this canonical basis, such that the eigenvectors of $Z_{1}$ and $Z_{2}$ will be linear combinations of these QNSV or, more generally, qu$n$its of the form (\ref{qnit}) for a fixed value of $n$.

For example, the qu1it orthonormal basis states $\big\{|1,-1\rangle, |1,0\rangle, |1,+1\rangle \big\}$ are represented by
\begin{equation*} \label{qutrit}
     |1,-1\rangle  = \left( \begin{array}{cc}
         1\\
         0\\
         0\\
    \end{array}\right), \quad |1,0\rangle = \left( \begin{array}{cc}
         0\\
         1\\
         0\\
    \end{array}\right),  \quad |1,0\rangle = \left( \begin{array}{cc}
         0\\
         0\\
         1\\
    \end{array}\right) \\
\end{equation*}
and span a 3D Hilbert space, which can be realized by a three-level quantum system, such as in the case of spin 1. However, as discussed above, by restricting $\mathbf{Z}$ to have a 3D Euclidean representation, not every transition can be accessed separately. In turn, in the description of the Gell-Mann matrices \cite{gell-mann}, these states can be referred to true qutrits. Indeed, we need to use the full set of SU(3) operators \cite{furey2018} acting between basis states to properly described a qutrit. We discuss this as an example further in the following by generalizing our $Z$-representation.

Since we have defined the eigenvectors of the $q$-number $Z_{3}$ as being the vectors of the canonical basis, it is obvious that its QNSV are normalized. For the other components, using the column-matrix representation of QNSV, and considering the first term as equal to $1$, by convention, their general norm is well-defined as
\begin{equation} \label{norm}
     \big| \big| |n,m  \rangle \big| \big| = \sqrt{\langle n,m | n,m \rangle} = 2^{n}\sqrt{\frac{(n+m)!(n-m)!}{(2n)!}}
\end{equation}

To satisfy the properties (\ref{property4}) to (\ref{property7}) of $\mathbf{Z}$, the canonical matrix representations of its components are given by
\begin{equation} \label{zirep}
    [Z_{1}] = \left( \begin{array}{cccccc}
         0 & a & 0 & 0 & \dots & 0\\
         a&0&b&0&\dots&0\\
         0&b&0&c&\dots&0\\
         \vdots&\vdots&\vdots&\vdots&\ddots&\vdots\\
         0&0&0&0&k&0\\
    \end{array}\right)
\end{equation}
\begin{equation} \label{zjrep}
    [Z_{2}] = \left(\begin{array}{cccccc}
         0 & -ia & 0 & 0 & \dots & 0\\
         ia&0&-ib&0&\dots&0\\
         0&ib&0&-ic&\dots&0\\
         \vdots&\vdots&\vdots&\vdots&\ddots&\vdots\\
         0&0&0&0&ik&0\\
    \end{array}\right)
\end{equation}
\begin{equation} \label{zkrep}
    [Z_{3}] = \left(\begin{array}{cccccc}
         n&0&0&0&\dots&0\\
         0&(n-1)&0&0&\dots&0\\ 
         0&0&(n-2)&0&\dots&0\\
         \vdots&\vdots&\vdots&\vdots&\ddots&\vdots\\
         0&0&0&0&\dots&-n\\
    \end{array}\right)
\end{equation}
Now, the $c$-numbers appearing in the matrix representations, $a, b, c, k \in \mathbb{R}$ and the general matrix elements are calculated by using the formulas 
\begin{equation} \label{termzi}
(Z_{1})_{rs} = \textstyle\frac{1}{2} (\delta_{r, s+1} + \delta_{r+1, s})\sqrt{(n+1)(r+s-1)-rs}
\end{equation}
\begin{equation} \label{termzj}
(Z_{2})_{rs} = \textstyle\frac{i}{2} (\delta_{r, s+1} - \delta_{r+1, s})\sqrt{(n+1)(r+s-1)-rs}
\end{equation}
\begin{equation} \label{termzk}
(Z_{3})_{rs}=(n+1-r)\delta_{r,s}=(n+1-s)\delta_{r,s}
\end{equation}
From Eqs. (\ref{termzi}) to (\ref{termzk}), $n$ is related to the $(2n+1)$ dimension of the representation and $r$ ($s$) indicates the row (column) of the corresponding matrix. These representations are useful if one is interested in constructing quantum logic gates of higher dimensions, since it is always possible to complement their symmetry group. In fact, different matrix representations can be constructed from the 3D representation \cite{das2003}.

The characteristic polynomials associated to this matrix representation of the integer $q$-numbers show that the $Z_k$ eigenvalues are also algebraic numbers. This is, by calculating the the determinant of $(Z - xI)$, we find polynomials with whole coefficients for each chosen $(2n+1)$-dimension. For example, we exhibit the cases 
\begin{equation}
     \begin{array}{cc}
          n=1: D(x) = -x^{3}+x\\
          n=2: D(x) = -x^{5}+5x^{3}-4x\\
          n=3: D(x) = -x^{7}+14x^{5}-49x^{3}+36x\\
     \end{array}
 \end{equation}
whose roots are classical integer numbers. In addition, for the case of half-integer representation, we find characteristic polynomials, but with rational coefficients, e.g.,
\begin{equation}
     \begin{array}{cc}
          n=1/2: D(x) = x^{2} - \frac{1}{4}\\
          
          n=3/2: D(x) = x^{4} - \frac{5}{2}x^{2} + \frac{9}{16}\\
          
          n=5/2: D(x) = x^6 - \frac{35}{4}x^{4} + \frac{259}{16}x^{2} - \frac{225}{64}\\
          
     \end{array}
 \end{equation}
For all parity spaces, we obtain the general formula for the characteristic polynomials
\begin{equation}
   D(x) =  \prod\limits_{k=-n}^{+n} (k-x)
\end{equation}
We reinforce that for a QNT formulation within the $Z$-representation, Hilbert spaces of odd dimensions appears to be of more interest, since it deals with both integer $q$-numbers and $c$-numbers. Nonetheless, spaces of even dimensions can be useful for general quantum computing protocols.

\section{A quantum mapping of QNSV between distinct subspaces}

As a consequence of this $Z$-representation, an integer $c$-number can belong to QNSV of different dimensions. It is also know that there exists a relationship between the dimension $d$ of the $Z$-space and a given value of a natural $c$-number $n$, i.e., $d=2n+1$. In this sense, there exists a subspace $\mathbb{V}_{min} \subset \mathcal{H}$ with $\text{dim}{(\mathbb{V}_{min})}=d_{min} = 2n_{min}+1$, such that $|n_{min},m\rangle \in \mathbb{V}_{min}$ is the QNSV with the smallest dimension associated to the eigenvalue $m$ for a specific $\mathbf{Z}$-component. Therefore, there are larger QNSV $|n,m\rangle \in \mathbb{V}$, with $\text{dim}(\mathbb{V})>\text{dim}(\mathbb{V}_{min})$, which also possess $m$ as an eigenvalue. By convention, we define the QNSV belonging to $\mathbb{V}_{min}$ as the `fundamental' QNSV in the $Z$-representation.

As previously discussed in Section III, the odd dimensions generate $\mathbb{Z}$ and even dimensions generate $\frac{1}{2}\mathbb{Z}^{*}$. Thus, all integer $c$-number generated in a dimension $d$ are also generated in a dimension $d+2$. This implies that for all $m \in \mathbb{Z}\cup\frac{1}{2}\mathbb{Z}^{*}$, there is a collection of QNSV $\{|n_{min}, m\rangle, |n_{min}+1, m\rangle, |n_{min}+2, m\rangle, ...\}$ containing $m$ as an eigenvalue. In this way, we propose a homomorphism $T_{p}$, $p=1, 2, 3$, between vector spaces that relate two QNSV of different dimensions, exhibiting the same eigenvalue. In this sense, we define $T_p$ as a $q$-mapping for the component $p$.

Let us consider $T_p$ as a homomorphism between $\mathbb{V}_{min}$ and $\mathbb{V'}$, with $\text{dim}(\mathbb{V'})=\text{dim}(\mathbb{V}_{min})+2$ to ensure the same parity between spaces:
\begin{equation}\label{quantummap}
    T_p|n_{min},m \rangle = |n_{min}+1, m \rangle
\end{equation}
For a space $\mathbb{V}$ with $\text{dim}(\mathbb{V}) = d_{min} + 2v$ ($v=1,2,3,\dots$), we find the homomorphism $T_p: \mathbb{V}_{min} \shortrightarrow \mathbb{V}$ as follows
\begin{equation}\label{multipleQM}
    T_{(n_{min}\shortrightarrow n)} = T_{(n-1 \shortrightarrow n)} \circ T_{(n-2 \shortrightarrow n-1)} \circ \dotsb \circ T_{(n_{min}\shortrightarrow n_{min}+1)}
\end{equation}
with the operation `$\circ$' being the composition between distinct $q$-mappings. We notice that there are $v-1$ compositions, indicating that knowing the fundamental QNSV $|n_{min},m \rangle$, it is possible to find any other $|n_{min}+v,m\rangle$ that satisfy Eq. (\ref{zk}) with the generalization given by Eq. (\ref{multipleQM}).

For each component $p=1,2,3$, we obtain a matrix representation in the integer QNSV basis for all the $q$-mappings $T_p$. Their general matrix elements are determined by 
\begin{equation} \label{T1}
    \big(T_{1}\big)_{rs}=\delta_{r,s}\prod \limits_{k=1}^{r-1}\sqrt{\frac{d-k}{d+2-k}} - \delta_{d+2-r,d-s}\prod \limits_{k=1}^{d-s}\sqrt{\frac{d-k}{d+2-k}}
\end{equation}
\begin{equation} \label{T2}
    \big(T_{2}\big)_{rs}=\delta_{r,s}\prod \limits_{k=1}^{r-1}\sqrt{\frac{d-k}{d+2-k}} + \delta_{d+2-r,d-s}\prod \limits_{k=1}^{d-s}\sqrt{\frac{d-k}{d+2-k}}
\end{equation}
\begin{equation} \label{T3}
    \big(T_{3}\big)_{rs}=\delta_{r+1,s}
\end{equation}
with $d=2n+1$.

For completeness, we define a new operator $R_p$ denoted quantum mapping operator (QMO), given by

\begin{equation} \label{retracted}
     R_p \equiv T_p^{\dagger}T_p
\end{equation}

Using the adjoint relation of Eq. (\ref{quantummap}), i.e., $\langle n, m|T^{\dagger}=\langle n+1, m|$, we obtain for Eq. (\ref{retracted})
\begin{equation} \label{Rprojection}
    \langle n, m|R_{p}|n, m\rangle = \langle n+1, m|n+1, m\rangle
\end{equation}
Since the right hand side of Eq. (\ref{Rprojection}) is larger than zero, $R_p|n, m\rangle$ must be a multiple of $|n, m\rangle$. Thus, we can establish the following eigenvalue equation

\begin{equation} \label{eigenReduced}
    R_{p} |n, m\rangle = r|n, m\rangle
\end{equation}
In this sense, all QNSV $|n,m\rangle$ are simutaneously eigenvectors of both operators $Z_p$ and $R_p$ ($p=1,2,3$).

Applying Eq. (\ref{T3}) in Eq. (\ref{retracted}), we obtain that $R_{3}=I$ that exhibits $r=1$ as its single eigenvalue. For $p=1$ or $p=2$, we use the general norm of $|n, m\rangle$ given by Eq. (\ref{norm}) to find all the eigenvalues:

\begin{equation} \label{Reigenvalues}
    r = \frac{\langle n+1, m|n+1, m\rangle}{\langle n, m|n, m\rangle}
\end{equation}
This equation implies that $r$ is a function of the dimension $d$ and the $c$-number $m$. For a given $d=2n+1$, we obtain
\begin{equation} \label{rfunction}
    r(m) = 1+ \frac{1}{d}-\frac{4m^2}{d(d+1)}
\end{equation}
Eq. (\ref{rfunction}) has the following meaning: i) $1+1/d$ is always a $c$-number of $R_p$ iff $d$ is odd, because the eigenvalue $0$ belongs to the set generated by $Z_p$ with odd dimension; ii) there is a degeneracy between the QNSV $|n, \pm m\rangle$, because $r(+m)=r(-m)$; iii) when $|m|=n=\frac{d-1}{2}$, we obtain $r=\frac{4}{d-1}$.

Finally, if we use the negative QMO, i.e., $-R_{p}$, it is possible to generate the eigenvalues $w \in \mathbb{Q}^{*}_{-}$. Therefore, the algebraic structure of QMO allows one obtains a subset $W \subseteq \mathbb{Q}^{*}$. At this point, it is important mentioning that in our proposed QNT all $c$-numbers appear to be displayed from an eigenvalue equation involving a specific $q$-number. This is, however, our main aim within this algebraic framework of a number theory. 

\section{The interplay between the $\mathbf{Z}$-representation and SU($n$)}
As proposed in Section III, the integer $q$-number representation leads to three components that are basis elements of the Lie algebra $\mathfrak{su}(n)$. In this sense, considering the quantum computing context, the QNSV associated to the $Z$-components do not represent true quantum information units, but number information basic units. We have previously given as an example the case of a qutrit, i.e., a qu$n$it for $n=1$ in $d=2n+1$. Indeed, a true qutrit requires the full set of SU(3) operators acting between basis states to be properly described as a quantum information entity. However, it is possible, by construction, to relate the $q$-number $\mathbf{Z}=(Z_{1}, Z_{2}, Z_{3})$ to SU($n$), in the same way that we can obtain all SU(3) matrices (the $3 \times 3$ Gell-Mann matrices $\big\{ \lambda_{i} \big\}$ in this case \cite{gell-mann}) from a 3D representation of $\mathbf{Z}$. 

As is well known for a qutrit, these matrices can be described in terms of eight $q$-number operators, which can be derived from the $\mathbf{Z}$-components; i.e, the operators $Z_{1}$, $Z_{2}$, $Z_{3}$, $Z^2_{1}$, $Z^2_{2}$, $\big\{Z_{1},Z_{2} \big\}$, $\big\{Z_{2},Z_{3} \big\}$, and $\big\{Z_{3},Z_{1} \big\}$, where $\big\{Z_{i},Z_{j} \big\}=Z_{i}Z_{j}+Z_{j}Z_{i}$ denotes the anticommutators with ($i,j=1,2,3)$. Hence, the decomposition of the Gell-Mann matrices in terms of qu1nit matrices is as follows:
\begin{equation} \label{gellmann}
\begin{split}
     \lambda_{1}  = \frac{1}{\sqrt{2}}(Z_{1} + \big\{Z_{3},Z_{1} \big\}), \quad \lambda_{2} = \frac{1}{\sqrt{2}}(Z_{2} + \big\{Z_{2},Z_{3} \big\})  \\
    \lambda_{3} =2I + \frac{1}{2}(Z_{3} -3Z^2_{1} - 3Z^2_{2}) , \quad \lambda_{4} = (Z^2_{1} - Z^2_{2})\\
    \lambda_{5} = \big\{Z_{1},Z_{2} \big\}, \quad \lambda_{6} = \frac{1}{\sqrt{2}}(Z_{1} - \big\{Z_{3},Z_{1} \big\}) \\
    \lambda_{7} = \frac{1}{\sqrt{2}}(Z_{2} - \big\{Z_{2},Z_{3} \big\}), \quad \lambda_{8} = \frac{1}{\sqrt{3}} \left(-2I + \frac{3}{2}(Z_{3} + Z^2_{1} + Z^2_{2})\right)
\end{split}
\end{equation}
Notice that the matrix representations of the $\mathbf{Z}$-components in Eqs. (\ref{gellmann}) are 3D:
\begin{equation} \label{eight-param}
    \begin{split}
        Z_{1} ={\frac{1}{\sqrt{2}}\left(\begin{array}{ccc}
             0&1&0  \\
             1&0&1 \\
             0&1&0\\
        \end{array}\right)}, \quad Z_{2}={\frac{1}{\sqrt{2}}\left(\begin{array}{ccc}
             0&-i&0  \\
             i&0&-i\\
             0&i&0\\
        \end{array}\right)}, \quad Z_{3}={\left(\begin{array}{ccc}
             1&0&0  \\
             0&0&0\\
             0&0&-1\\
        \end{array}\right)},\\
        \{Z_{1},Z_{2}\}=\left(\begin{array}{ccc}
             0&0&-i  \\
             0&0&0\\
             i&0&0\\
        \end{array}\right), \quad \{Z_{2},Z_{3}\} = \frac{1}{\sqrt{2}}\left(\begin{array}{ccc}
             0&-i&0  \\
             i&0&-i\\
             0&-i&0\\
        \end{array}\right)\\
        \{Z_{3},Z_{1}\}=\frac{1}{\sqrt{2}}\left(\begin{array}{ccc}
             0&1&0  \\
             1&0&-1\\
             0&-1&0\\
        \end{array}\right), \quad Z_{1}^{2} = \left(\begin{array}{ccc}
             \frac{1}{2}&0&\frac{1}{2}  \\
             0&1&0\\
             \frac{1}{2}&0&\frac{1}{2}\\
        \end{array}\right), \quad Z_{2}^{2}=\left(\begin{array}{ccc}
            \frac{1}{2}&0&-\frac{1}{2}  \\
             0&1&0\\
             -\frac{1}{2}&0&\frac{1}{2}\\
        \end{array}\right)
    \end{split}
\end{equation}
With this representation, it is possible to search quantum computing protocols for a single qutrit \cite{dogra2018} or compute entanglement starting with two qutrits on a proper Hilbert space \cite{bertlmann2008}. 

The Gell-Mann matrices described by Eqs. (\ref{gellmann}) satisfy the following commutation relations
\begin{equation} \label{comm-gell}
    [\lambda_{a}, \lambda_{b}] = if_{abc}\lambda_{c}, \quad (a,b,c = 1,2,3,...,8)
\end{equation}
with an implicit sum over $c$, and $f_{abc}$ being the structure constants totally antisymmetric by interchanging any pair of indices, defining an $\mathfrak{su}(3)$ algebra. Furthermore, these matrices have the following properties
\begin{equation} \label{prop-gell}
   \textup{Tr}(\lambda_{a} \lambda_{b}) = 2\delta_{ab}, \quad \big\{\lambda_{a},\lambda_{b}\big\} = 2d_{abc}\lambda_{c} + \textstyle\frac{4}{3}\delta_{ab}I
\end{equation}
and
\begin{equation} \label{sub-gell}
    f_{abc} = -\textstyle\frac{1}{4}i\textup{Tr}(\lambda_{a}[\lambda_{b}, \lambda_{c}]), \quad d_{abc} = \frac{1}{4}\textup{Tr} (\lambda_{a} \big\{\lambda_{b},\lambda_{c}\big\})
\end{equation}
where the elements $d_{abc}$ are totally symmetric by interchanging any pair of indices.

Using these properties, the density matrix of a pure qutrit in the 3D representation, $\rho_Q$, is given by \cite{mendas2006}
\begin{equation} \label{qut-dens}
    \rho_{Q} = \frac{1}{3}(I + \sqrt{3}\textbf{b} \cdot \boldsymbol{\lambda})
\end{equation}
where \textbf{b} is related to $\textbf{a} \equiv (a_1, a_2,...,a_8)$, the generalized Bloch vector with eight real parameters, via $\textbf{b} \equiv \sqrt{3}\textbf{a}$ and $\boldsymbol{\lambda} \equiv (\lambda_1, \lambda_2,...,\lambda_8)$. It is worth remembering here that the real parameters in the vector \textbf{a} are related to average values of the general $q$-numbers given in Eq. (\ref{eight-param}) and a physical realization is given by the case of spin 1.

It is also possible to construct the quadratic Casimir operator in the representation of $\mathfrak{su}(3)$. From the explicit forms \cite{gell-mann} of the structure constants $f_{abc}$, one defines the Cartan-Killing metric tensor, and its inverse as well
\begin{equation} \label{cartan}
    g_{ab} = f_{acd}f_{bcd} = 3\delta_{ab}, \quad g^{ab} = \textstyle\frac{1}{3}\delta^{ab} 
\end{equation}
Thus, the quadratic Casimir operator is constructed
\begin{equation} \label{casimir}
    C_2 = \textstyle\frac{3}{4}g^{ab}\lambda_a\lambda_b = \frac{1}{4}\sum_{a}(\lambda_a)^2 = \frac{4}{3}I
\end{equation}
Note that $C_2$ can be defined for all $d$-dimensional irreducible $Z$-representations of $\mathfrak{su}(3)$. For this case, we denote the corresponding traceless Hermitian generators by $\mathcal{Z}_a$, with the normalization of the matrix generators in the defining $\mathfrak{su}(3)$ representation fixed by $\textup{Tr}(\mathcal{Z}_a\mathcal{Z}_b) = \frac{1}{2}\delta_{ab}$. Thus, for any irreducible representation $\mathcal{Z}$ of $\mathfrak{su}(3)$ the quadratic Casimir operator is defined in the form
\begin{equation} \label{casimir-n}
    C_2 = 3g^{ab}\mathcal{Z}_a \mathcal{Z}_b = \sum_{a}(\mathcal{Z}_a)^2 = c_{2Z}(I)_{d \times d}
\end{equation}
where the coefficient $c_{2Z}$ indicates that $C_2$ is a multiple of the $d$-dimensional identity. Thus, using Eq. (\ref{comm-gell}) for the $Z$-representation, we obtain
\begin{equation} \label{casimir-comm}
     [\mathcal{Z}_a, C_2] = 0, \quad (a = 1,2,3,...,8)
\end{equation}

In fact, to properly apply this construction for higher dimensions, we need to employ the generalized Gell-Mann matrices (GGM) to describe qudits \cite{bertlmann2008}; i.e, the $d^{2}-1$ matrices, with $d > 3$, generating the Lie algebra associated to the special unitary group SU($n$), $n \ge{2}$. Note that in the $Z$-representation $n \rightarrow 2n+1$, the dimension of the considered Hilbert subspace. Following Ref. \cite{bertlmann2008}, the GGM are defined as three different types of matrices, i.e., symmetric, antisymmetric, and diagonal. Also, by definition, all GGM are Hermitian and traceless.

For describing the $\mathcal{Z}$-representation, it is more convenient to define $d^2$ traceless $d \times d$ matrices of the form
\begin{equation} \label{ggm}
     (\mathcal{Z}_{l}^{k})_{ij} = \delta_{li}\delta_{kj} - \frac{1}{d}\delta_{kl}\delta_{ij}, \quad (i,j,k,l = 1,2,3,...,d)
\end{equation}
where $\mathcal{Z}_{l}^{k} = f_{lk}(Z_1,Z_2,Z_3)$. In this construction, 
\begin{equation} \label{ggm-sum}
    \sum_{l}(\mathcal{Z}_{l}^{l}) = 0
\end{equation}
indicating that of the  $d^2$ matrices, $\mathcal{Z}_{l}^{k}$, only $d^{2}-1$ are independent, which can be employed to generate the $\mathfrak{su}(n)$ Lie algebra. In this way, one obtains the corresponding commutation relations
\begin{equation} \label{comm-ggm}
    [\mathcal{Z}_{l}^{k}, \mathcal{Z}_{r}^{s}] = \delta_{r}^{k}\mathcal{Z}_{l}^{s} - \delta_{l}^{s}\mathcal{Z}_{r}^{k}
\end{equation}
where the matrices $\mathcal{Z}_{l}^{k}$ satisfy
\begin{equation} \label{herm-ggm}
   (\mathcal{Z}_{l}^{k})^{\dagger} = \mathcal{Z}_{k}^{l}
\end{equation}
Therefore, by using proper linear combinations, it is possible to use these auxiliary matrices to construct $d^2-1$ traceless $d \times d$ Hermitian matrices, which can be used to describe general quantum gates. By construction, all of them are also considered here as being general $q$-numbers in QNT.

\section{Final remarks}

In the present work, we employ the algebraic structure of quantum mechanics to extend the abstract concept of number. Using the Dirac's prescription, we define pure quantum number operators ($q$-numbers) of a Hilbert space that generate classical numbers ($c$-numbers) belonging to discrete Euclidean spaces, in terms of eigenvalue problems. These are (i) a 2-component natural $q$-number $\textbf{N}$, satisfying a Heisenberg-Dirac algebra, which is also a Lie algebra, and (ii) a 3-component integer $q$-number $\textbf{Z}$, obeying a Lie algebra structure. In the context of the present QNT, a probabilistic interpretation of number naturally emerges from the algebraic representation, extending the concept of number information. 

We interpret the eigenvectors of the $q$-numbers as QNSV, which form multidimensional orthonormal basis sets useful to describe state-vector superpositions defined as qu$n$its. The QNSV of different dimensions can be associated to the same integer $c$-number via a quantum mapping operation that relates distinct Hilbert subspaces. Furthermore, its structure can generate a subset $W \subseteq \mathbb{Q}^{*}$, the field of non-zero rationals. Mathematical operations in QNT can be related to quantum computing and enable understanding nontrivial computations in higher dimensions than qubits and qutrits. This is indeed an interesting and general description to improve the `number information notion', which can lead to new insights in both applied mathematics and quantum computing/information.

\section*{Author contributions}
All authors contributed equally to this work. Especially, Section IV was developed by LD.

\section*{Acknowledgments}
We thank the Brazilian agencies Conselho Nacional de Desenvolvimento Cient{\'i}fico e Tecnol{\'o}gico (CNPq) and Coordena{\c c}{\~a}o de Aperfei{\c c}oamento de Pessoal de N{\'i}vel Superior - Brasil (CAPES) - Finance Code 001. 

\section*{DATA AVAILABILITY}
Data sharing are available on request from the authors.

\bibliography{ref}
\end{document}